# T-Fusion Net: A Novel Deep Neural Network Augmented with Multiple Localizations based Spatial Attention Mechanisms for Covid-19 Detection

Susmita Ghosh[1], Abhiroop Chatterjee[1]

[1] Jadavpur University, Kolkata, India

**Abstract.** In recent years, deep neural networks are yielding better performance in image classification tasks. However, the increasing complexity of datasets and the demand for improved performance necessitate the exploration of innovative techniques. The present work proposes a new deep neural network (called as, *T-Fusion Net*) that augments multiple localizations based spatial attention. This attention mechanism allows the network to focus on relevant image regions, improving its discriminative power. A homogeneous ensemble of the said network is further used to enhance image classification accuracy. For ensembling, the proposed approach considers multiple instances of individual T-Fusion Net. The model incorporates fuzzy max fusion to merge the outputs of individual nets. The fusion process is optimized through a carefully chosen parameter to strike a balance on the contributions of the individual models. Experimental evaluations on benchmark Covid-19 (SARS-CoV-2 CT scan) dataset demonstrate the effectiveness of the proposed T-Fusion Net as well as its ensemble. The proposed T-Fusion Net and the homogeneous ensemble model exhibit better performance, as compared to other state-of-the-art methods, achieving accuracy of 97.59% and 98.4%, respectively.

## 1 Introduction

Deep learning models (Fig. 1) are achieving greater success in the field of computer vision, especially in image classification tasks. Convolutional Neural Networks (CNNs) have demonstrated remarkable success in extracting discriminative features from images, enabling better classification. However, as datasets become more complex and diverse, achieving further improvements in classification performance remains a challenge. Researchers are continuously exploring innovative techniques to enhance the capabilities of image classification models. One area of focus is the integration of attention mechanisms into deep learning architectures and finding the suitable position for incorporating them. Attention mechanisms aim to enhance the discriminative power of models by selectively focusing on relevant regions within an image. Such mechanisms allow the models to attend to important features and suppress irrelevant or noisy information, leading to improved classification accuracy. Spatial attention, in particular, has gained popularity for its ability to capture fine-grained details from images.

In this context, the present work proposes a novel deep neural network that integrates *MLSAM*, multiple localizations based spatial attention mechanisms. The augmented network is termed as T-Fusion Net. A homogeneous ensemble model of this T-Fusion Net is further used that leverages the strengths of multiple individual model objects, each incorporating spatial attention with different kernel sizes. The fusion of outputs from each individual model is achieved through fuzzy max fusion.

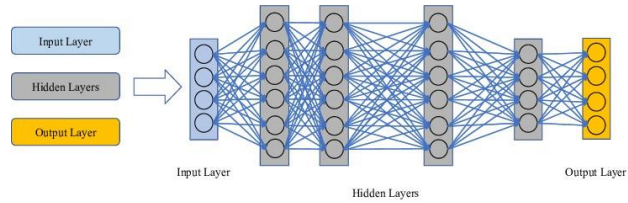

**Fig. 1.** General representation of deep neural net [7]

The effectiveness of our proposed approach is evaluated through experiments on benchmark Covid-19 detection dataset: SARS-CoV-2 CT scan [4]. Performance comparison with several other deep neural network models is done. It is seen that the performance of the proposed T-Fusion Net augmented with spatial attention and thereafter ensembling it with fuzzy max fusion is showcasing its superiority in terms of accuracy and other important parameters.

The rest of the paper is organized as follows: Section 2 provides a review of related literature in the field of image classification using deep learning. Section 3 presents the methodology, including a detailed description of the proposed multiple localizations based spatial attention block, its integration within the T-Fusion Net architecture, and the ensemble through fuzzy max fusion. Section 4 discusses the experimental setup mentioning dataset used, evaluation metrics considered, image preprocessing and details of parameters taken. Analysis of results has been put in Section 5. Finally, Section 6 concludes the paper.

## 2    Related Research

Zhang et al. [1] proposed a hybrid attention method that combined both spatial and channel attention mechanisms. Attention mechanisms in deep learning models allow the network to focus on specific regions or channels of input data that are most relevant for making predictions. Spatial attention helps the model concentrate on important spatial regions, while channel attention emphasizes important channels in the feature maps. The approach demonstrated significant performance gains compared to conventional methods, making it a state-of-the-art technique at the time of its publication. Huang et al. [2] introduced a fuzzy fusion technique as an effective way to combine outputs from individual models. Ensemble methods, such as model fusion, are commonly used to boost the performance of machine learning models. In the context of image classification, multiple models may produce different predictions for the same image, and combining their outputs can lead to improved accuracy and robustness. Zheng et al. [3] proposed a hierarchical fusion approach that combined multiple levels of features for image classification. In deep learning models, features are hierarchically learned at different layers of the network.

In line with these advancements, our research presents a novel deep neural network model (termed as, T-Fusion Net) that integrates a new spatial attention method. By incorporating attention mechanisms into individual models, we aim to capture diverse and discriminative image features. Thereafter, using fuzzy max fusion, our model optimally combines the strengths of individual T-Fusion Net, leading to improved classification accuracy.

## 3  Proposed Methodology

In the present work, the architecture of the proposed model is designed using Convolutional Neural Networks (CNNs) to extract meaningful features from images. Spatial attention is incorporated by adding convolutional layers with different kernel sizes. We call it multiple localizations. The outputs of these attention-enhanced convolutional layers are concatenated to capture discriminative features from images. Since the proposed network model looks like the English alphabet "T", we termed it as *T-Fusion Net*. Additionally, we make an ensemble model with a fuzzy max function to further improve the performance of the T-Fusion Net. Each of the modules of the T-Fusion Net is described below in detail.

### 3.1  Multiple Localizations based Spatial Attention

The proposed multiple localizations based attention mechanism (*MLSAM*) aims to enhance the receptive power of the model by selectively focusing on important regions within the input feature map.

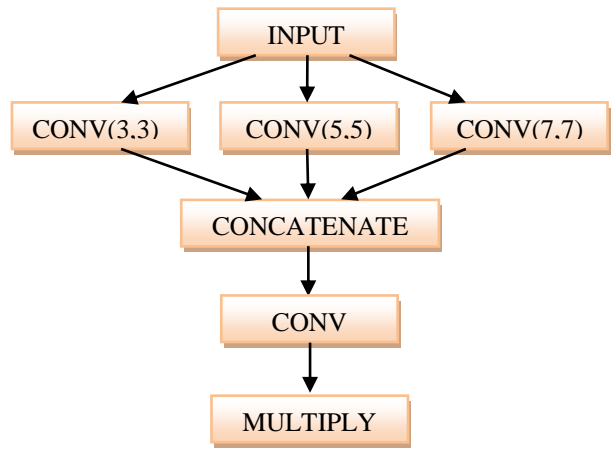

**Fig. 2.** MLSAM: Multiple Localization based Spatial Attention Mechanism.

The MLSAM module in Fig 2. takes an input feature map and applies convolutional operations to capture local and global patterns at different scales by varying kernel sizes. It consists of three parallel branches, each performing convolution with different kernel sizes: 3x3, 5x5, and 7x7. These branches generate feature maps with 4 channels each, resulting in a concatenated feature map with a total of 12 channels. The concatenation of the feature maps is performed along the channel axis, allowing the model to capture diverse information from different kernel sizes. This step is crucial for acquiring multi-scale feature extraction. Subsequently, a convolutional layer with a 3x3 kernel size is applied to the concatenated feature map. This layer reduces the number of channels to one using the sigmoid activation function. The resultant output is a spatial attention map that represents the importance of different spatial locations within the input feature map. Finally, the spatial attention map is combined with the input feature map through element-wise multiplication. This operation selectively amplifies informative regions while suppressing less relevant ones.

### 3.2   Architecture of T-Fusion Net

The architecture of the T-Fusion Net, as shown in Fig. 3, consists of several convolutional layers, which are fundamental building blocks for extracting hierarchical features from input images. The architecture follows a sequential pattern, progressively transforming the given input images to higher-level representations.

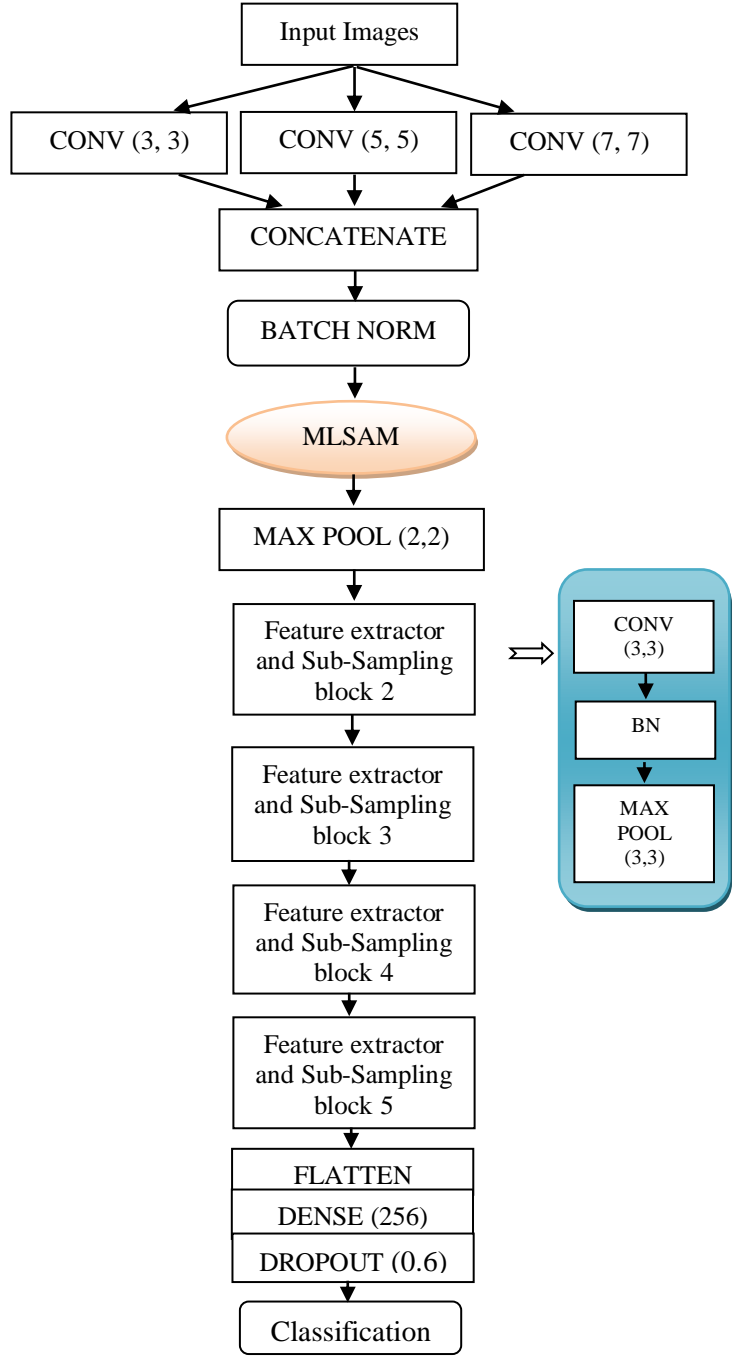

**Fig. 3.** T-Fusion net architecture.

The details of each of the components of the T-Fusion Net are described below in brief.

Input Layer: The input layer takes in images with a shape of (224, 224, 3), representing the width, height, and three color channels (R, G, B).

Convolutional Layers: The initial convolutional layer applies a set of filters to the input image, extracting local patterns and low-level features. Three separate convolutional layers with different kernel sizes (3x3, 5x5, 7x7) are applied in parallel, capturing information at different scales. Each convolutional layer performs element-wise multiplications followed by element-wise additions to create activation maps.

Concatenation: The outputs of the three parallel convolutional layers are concatenated along the channel axis. The purpose of concatenation is to capture diverse features at multiple scales, enabling the model to learn a richer representation of the input data.

Batch Normalization: Batch normalization is applied to normalize the activations of the concatenated feature maps, improving the convergence of the model.

MLSAM: Multiple Localizations based Spatial Attention Mechanism (MLSAM) is applied to the batch-normalized feature maps. As stated, this attention mechanism selectively emphasizes important spatial regions while suppressing irrelevant or noisy information thereby enhancing discriminative power. In the present work, the MLSAM has been introduced in the Feature extractor and subsampling Block 1 (in between Batch normalization and Max pooling) (please see Fig. 3).

Max Pooling: Max pooling is performed on the spatially attended feature maps to downsample the dimensions and reduce computational complexity.

Convolutional Layers and Pooling: Four additional convolutional layers and max pooling operations are applied sequentially to capture higher-level abstract features. These are represented in Fig. 3 as "Feature extractor and Sub-sampling block". On these, basically convolution, batch normalization and max pooling operations take place (as shown in enclosed blue block in Fig. 3). These layers gradually increase the receptive field, allowing the model to learn more complex patterns and relationships exist in the image data.

Flattening: The final feature maps are flattened into a 1-dimensional vector to serve as input for the subsequent fully connected layers.

Fully Connected Layers (Dense): The flattened feature vector is connected to a fully connected layer, which performs non-linear transformations to learn class-specific representations.

Dropout Regularization: Dropout regularization is applied to the fully connected layers to prevent overfitting. A dropout of 60% has been applied in this work.

Output (Softmax) Layer: The output layer consists of two nodes with softmax activation function, representing the probabilities of the given input image belonging to each of the two classes (Covid/ non-Covid). The class with higher probability is selected as the predicted class.

### 3.2   Ensemble Model

For a given input image, the probability values obtained from the softmax layer are treated as the membership values of belongingness to each of the classes. To perform the fusion of the outputs of individual T-Fusion Nets, a fuzzy max fusion method has been applied that combines the outputs of individual nets. The homogeneous ensemble (Fig. 4) computes the maximum value across the individual model outputs. The fused output is obtained by balancing the output with α, adding ϵ, and further adding a bias B. Finally, the technique constructs the ensemble model, specifying the input and output layers.

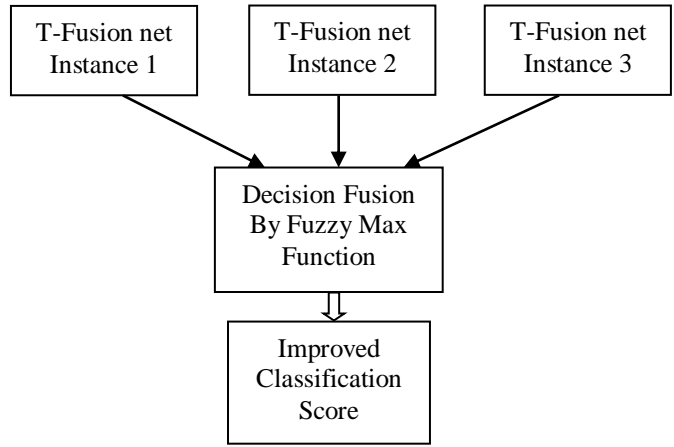

**Fig. 4.** Proposed ensemble model.

This ensemble technique utilizes the fuzzy max fusion to combine the predictions from multiple individual models. $F_o$, the fused output is written as:

$$F_o = \alpha * M_o + \varepsilon + B \quad (1)$$

where, $M_o$ is the maximum value obtained from all the three individual nets, α determines the balance between the individual model outputs and the fused output. ε is a small constant introduced to avoid division by zero, and a bias $B$ is applied to introduce some offset and ensure numerical stability.

## 4   Experimental Setup

To assess the effectiveness of the proposed T-Fusion Net and thereafter its homogeneous ensemble, experiment has been conducted using the SARS-CoV-2 CT scan dataset. Details of this experimental setup have been described below in detail.

### 4.1   Dataset Used

SARS-CoV-2 CT scan dataset [4] is used. It consists of 1252 images of COVID-19 and 1230 images of Non COVID-19 cases (Table 1). Sample images from both the classes have been shown in Fig. 5.

**Table 1.** Images taken from SARS-CoV-2 CT scan dataset [4].

| Types of Classes | Number of Images |
|---|---|
| COVID-19 | 1252 |
| NON COVID-19 | 1230 |
| **Total Images** | **2482** |

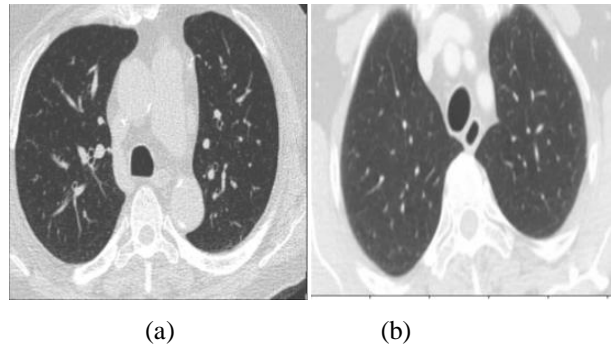

(a)  (b)

**Fig. 5.** Images taken from SARS-CoV-2 CT scan dataset. (a) Covid-19, (b) Non Covid-19

### 4.2   Image Preprocessing

Each input image is resized to a dimension of 224x224 pixels and normalized to [0,1] by dividing the pixel values by 255.

### 4.3   Performance Metrics Considered

The models' performance is evaluated using metrics such as accuracy, loss values. We have also considered other performance evaluation criteria e.g., Precision, Recall, F1-score, Top-1% error. Confusion matrix and IoU curve have also been considered.

### 4.4   Parameters Taken

Table 2 shows the parameters and their corresponding values used for our experimentation. Also the values of $\alpha$, $\varepsilon$ and B are set to 0.8, 0.0001 and 20 respectively (though we have experimented with various values of these parameters).

**TABLE 2.** Experimental setup.

| Parameters | Values |
|---|---|
| Learning Rate | 0.0001 |
| Batch Size | 16 |
| Max Epochs | 50 |
| Optimizer | Adam |
| Loss Function | Categorical Cross-entropy |

### 4.5   Model Training

The dataset is split into training and testing sets. The split is performed with a test size of 20% and stratified sampling to maintain class balance. During training, the models' weights are updated using backpropagation and gradient descent to minimize the loss function.

## 5   Analysis of Results

To evaluate the effectiveness of the proposed T-Fusion Net and also its ensemble with fuzzy fusion, various performance metrics are used, as mentioned earlier. A total of 20 simulations have been performed and the average values obtained for each of the metrics are depicted into Table. 3. From the table it is noticed that the results are promising in nature in terms of various performance indices, yielding 98.4% accuracy

8    Susmita Ghosh et al.

for the ensemble of T-Fusion Nets. Experimentation was done on NVIDIA A100 tensor core GPU.

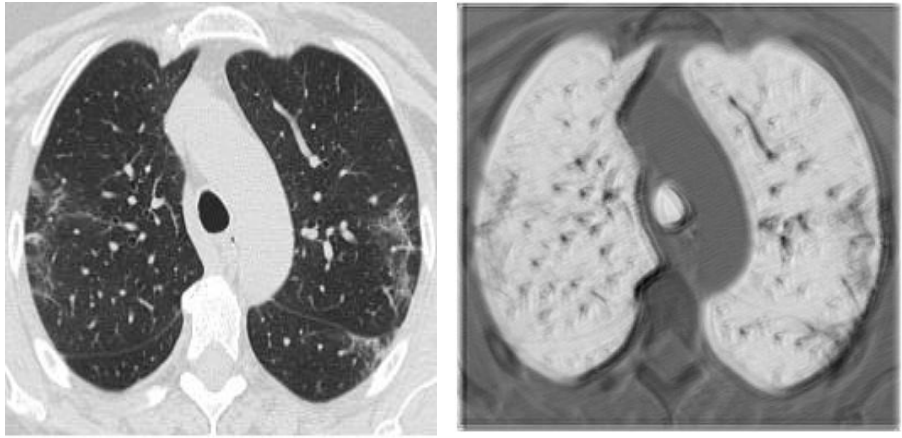

**Fig. 6.** Original input image is represented in left while the visualization of the intermediate feature representation after the proposed MLSAM module in T-Fusion Net is shown in right.

As described earlier, in the present paper, we propose a new soft segmentation approach called (MLSAM) for image classification tasks. MLSAM aims to enhance the interpretability and accuracy of soft segmentation models. This is achieved by incorporating multiple localizations based spatial attention mechanisms, which allow the model to selectively focus on different regions of the image at various levels of granularity. Fig. 6 shows how our MLSAM module segments the original image.

**TABLE 3.** Results obtained using SARS-CoV-2 CT scan dataset

| Metrics | Ensemble Model (rounded) |
|---|---|
| Precision | 0.98 |
| Recall | 0.98 |
| F1-score | 0.98 |
| Accuracy (%) | 98.0 |
| Top-1 error (%) | 2.0 |

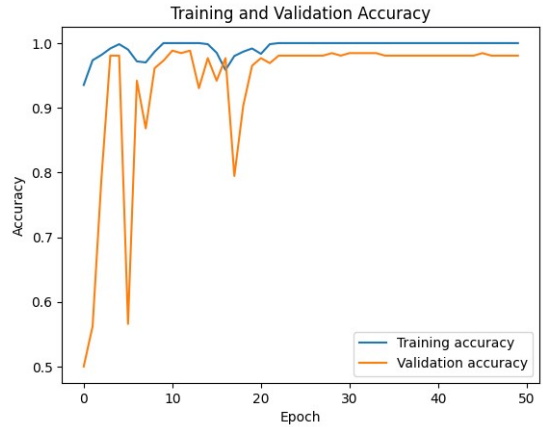

**Fig. 7.** Variation of training and validation accuracy with epochs for T-Fusion net.

Variations of training & validation accuracy and training & validation loss over epochs for the T-Fusion Net are shown in Figs. 7 and 8, respectively. From Fig. 7 it is seen that both training and validation accuracy increase over epochs. This suggests that the model is generalizing the data well and can make accurate predictions. As training progresses, the validation and training accuracies continue to steady down. Likewise, as expected, loss values are also getting stabilized over epochs (Fig. 8).

Fig. 9 shows the IoU bar plots for two different classes. IoU is a commonly used metric in tasks such as object detection and image segmentation. It measures the overlap between the predicted and true positive classes. In this case, the IoU values are 0.9538 and 0.9522 for Covid-19 and Non Covid-19, respectively. This indicates a high degree of overlap between the predicted positive class and the true positive class for the two respective classes.

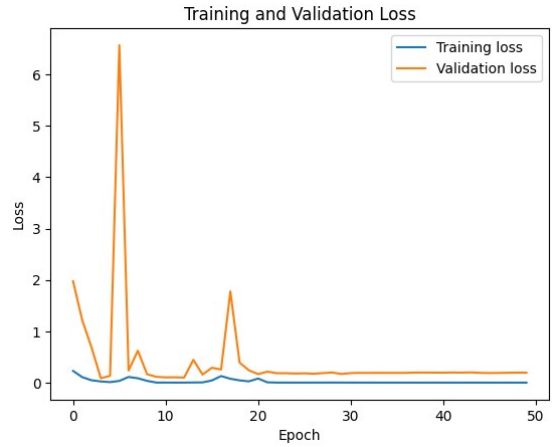

**Fig. 8.** Variation of training and validation accuracy with epochs for T-Fusion net +  MLSAM.

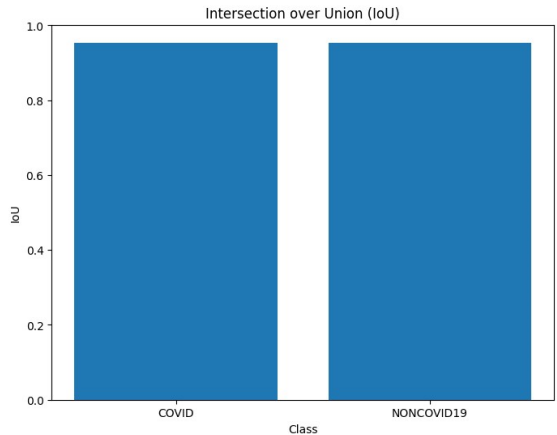

**Fig. 9:** IoU (Intersection over Union) curves for Covid-19 and Non Covid-19 classes.

The confusion matrix obtained using the proposed T-Fusion Net (augmented with MLSAM) is shown in Fig. 10. The entries in the matrix depict the best result obtained out of 20 simulations. The resultant matrix indicates that the proposed model is performing very well for Covid-19 detection.

To establish the efficacy of our proposed network, performance of the T-Fusion Net and also its ensemble have been compared with four pretrained (on ImageNet dataset) deep learning models (AlexNet, VGG-16, VGG-19, and DenseNet201) and also with an explainable deep learning approach. The corresponding accuracy values are shown in Table 4. The last 3 rows of the table (marked as bold) depict the results obtained through our proposed network and averaged over 20 simulations. In this connection it is to be noted that, the proposed network is trained from scratch, and not pretrained. This table confirms the superiority of T-Fusion Net for Covid-19 detection, even without ensemble.

**TABLE 4.** Classification performance of the proposed and the state-of-the-art models for the SARS-CoV-2 CT scan dataset.

| Methods | Accuracy (%) |
|---|---|
| AlexNet (Pretrained on Imagenet) [5] | 93.71 |
| VGG-16 (Pretrained on Imagenet) [5] | 94.62 |
| VGG-19 (Pretrained on Imagenet) [5] | 93.56 |
| DenseNet201based deep TL [6] | 96.25 |
| **T-Fusion net (baseline, no MLSAM)** | **96.59** |
| **T-Fusion net (with MLSAM)** | **97.59** |
| **Ensemble of T-Fusion net (with MLSAM)** | **98.40** |

We have introduced the T-Fusion Net architecture and explored its performance with and without the MLSAM. The proposed deep neural network not only achieves an accuracy of 97.59% (better than other existing methods) but also has less number of parameters (4,221,947) compared to other state-of-the-art neural networks. The Ensemble T-Fusion Net with MLSAM achieved the highest accuracy of 98.40%, outperforming other models and approaches.

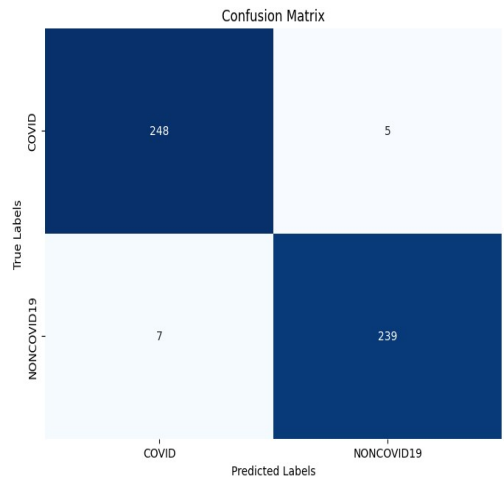

**Fig. 10:** Confusion matrix for Covid-19 detection for the proposed T-Fusion net (with MLSAM)

## 6  Conclusion and Future Works

The present work introduces a new deep neural network called *T-Fusion Net* and incorporates a novel spatial attention mechanism (termed MLSAM). An ensemble of such nets with fuzzy max fusion is also employed. The models were trained to extract relevant features and classify images into Covid-19 and Non Covid-19 categories. The evaluation of these models using various metrics demonstrated their effectiveness in identifying COVID-19 cases. The results obtained from this research (with accuracy of 97.50% and 98.4%, respectively for individual T-Fusion Net and its ensemble) offer promising prospects for use of multiple localizations based spatial attention in Covid-19 diagnosis.

However, further research and refinement of these models are needed to explore it in diverse domains. Future studies should focus on expanding the dataset to include a wider range of Covid-19 cases, including different imaging modalities and disease stages. Additionally, the models could benefit from fine-tuning and optimization to enhance their performance and generalization capabilities.

## Acknowledgement

A part of this work has been supported by the IDEAS - Institute of Data Engineering, Analytics and Science Foundation, The Technology Innovation Hub at the Indian Statistical Institute, Kolkata through sanctioning a Project No /ISI/TIH/2022/55/ dtd. September 13, 2022